\documentclass[twocolumn,aps,prl,showpacs,superscriptaddress,preprintnumbers,floatfix,nofootinbib,longbibliography]{revtex4-1}
\usepackage{multirow}
\usepackage{hhline}
\usepackage{bbm}
\usepackage{epsfig}
\usepackage{graphics}  
\usepackage{graphicx}  
\usepackage{dcolumn}   
\usepackage{bm}        
\usepackage{amssymb}   
\usepackage{amsmath}   
\usepackage{amsfonts}
\usepackage[percent]{overpic}
\usepackage{natbib}
\usepackage[colorlinks,linkcolor=blue,urlcolor=blue,citecolor=blue]{hyperref}
\usepackage[usenames]{color}
\usepackage{CJK}

\newcommand{\pT} {p_{\mathrm{T}}}
\newcommand{\lr}[1]{\left\langle #1\right\rangle}

\newcommand{\oo}{$^{16}$O+$^{16}$O }

\newcommand{\abini}{{\textit{ab-initio} }}

\newcommand{\trento}{T$\mathrel{\protect\raisebox{-2.1pt}{R}}$ENTo}

\begin{document}

\title{\textit{Ab-initio} nucleon-nucleon correlations and their impact on high energy \oo{}collisions}
\newcommand{\sbu}{Department of Chemistry, Stony Brook University, Stony Brook, NY 11794, USA}
\newcommand{\bnl}{Physics Department, Brookhaven National Laboratory, Upton, NY 11976, USA}
\newcommand{\moe}{Key Laboratory of Nuclear Physics and Ion-beam Application (MOE), and Institute of Modern Physics, Fudan
University, Shanghai 200433, China}
\newcommand{\fudan}{Shanghai Research Center for Theoretical Nuclear Physics, NSFC and Fudan University, Shanghai 200438, China}
\newcommand{\heid}{Institut f$\ddot{u}$r Theoretische Physik, Universit$\ddot{a}$t Heidelberg, Philosophenweg 16, 69120 Heidelberg, Germany}
\author{\small Chunjian Zhang}\email{chunjianzhang@fudan.edu.cn}\affiliation{\moe}\affiliation{\fudan}\affiliation{\sbu}
\author{\small Jinhui Chen}\email{chenjinhui@fudan.edu.cn}
\affiliation{\moe}\affiliation{\fudan}
\author{\small Giuliano Giacalone}\email{giacalone@thphys.uni-heidelberg.de}
\affiliation{\heid}
\author{\small Shengli Huang}\email{shengli.huang@stonybrook.edu}\affiliation{\sbu}
\author{\small Jiangyong Jia}\email{jiangyong.jia@stonybrook.edu}
\affiliation{\sbu}\affiliation{\bnl}
\author{\small Yu-Gang Ma}\email{mayugang@fudan.edu.cn}
\affiliation{\moe}\affiliation{\fudan}

\begin{abstract}
Investigating nucleon-nucleon correlations inherent to the strong nuclear force is one of the core goals in nuclear physics research. We showcase the unique opportunities offered by collisions of $^{16}$O nuclei at high-energy facilities to reveal detailed many-body properties of the nuclear ground state. We interface existing knowledge about the geometry of $^{16}$O coming from \textit{ab-initio} calculations of nuclear structure with transport simulations of high-energy $^{16}$O+$^{16}$O collisions. Bulk observables in these processes, such as the elliptic flow or the fluctuations of the mean transverse momentum, are found to depend significantly on the input nuclear model and to be sensitive to realistic clustering and short-range repulsive correlations, effectively opening a new avenue to probe these features experimentally. This finding demonstrates collisions of oxygen nuclei as a tool to elucidate initial conditions of small collision systems while fostering connections with effective field theories of nuclei rooted in quantum chromodynamics (QCD).
\end{abstract}
\maketitle

{\it Introduction}-- Unveiling the structure and strongly-correlated nature of atomic nuclei across the nuclide chart has been the subject of sustained interest in fundamental physics~\cite{RevModPhys.70.743,VONOERTZEN200643,RevModPhys.90.035004,RevModPhys.81.1773,Kondo:2023lty}. Information on atomic nuclei is predominantly obtained through low-energy experiments, such as elastic electron scattering and Coulomb excitation~\cite{Cline:1986ik,Yang:2022wbl}, in combination with advanced phenomenological models~\cite{Yang:2022wbl}. The simplest model of an atomic nucleus is a collection of non-interacting fermions in space following, e.g., a three-parameter Fermi (3pF) density~\cite{DEVRIES1987495},
\begin{equation}\label{eq0}\begin{split}
\rho(r) \propto  \frac{ 1+w\left(r^2/R^2\right) }{1+e^{(r-R)/a_0}},
\end{split}
\end{equation}
where $R$ represents the half-width radius, $a_0$ specifies the surface diffuseness, while $w$ quantifies the central density depletion. For large enough nuclei, knowledge of the function $\rho(r)$ (supplemented with spatial deformation) is to a great extent enough to capture the main properties of their structure and systematically explain the related experimental signatures~\cite{Bender:2003jk}. 

For light or intermediate-mass ($A\approx20$) nuclei, a mean-field description breaks down and two-body nucleon-nucleon ($NN$) correlations acquire prominent importance~\cite{RevModPhys.90.035004,Cruz-Torres:2019fum}. Modern \textit{ab-initio} approaches to the nuclear many-body problem~\cite{Hergert:2020bxy,Ekstrom:2022yea} rooted in effective field theories of low-energy QCD~\cite{Hammer:2019poc,Piarulli:2019cqu} are nowadays capable of addressing from first principles deformed intermediate-mass species and the emergence of clustering correlations therein~\cite{Hagen:2013nca,Hergert:2015awm,Soma:2019bso,Stroberg:2019bch,Tichai:2020dna,Otsuka:2022bcf,Chen:2023mel,Ekstrom:2023nhc,Arthuis:2024mnl,Giacalone:2024luz,Sun:2024iht}. These models are typically benchmarked against experimental data on binding energies, charge radii, or spectroscopic information which nonetheless probe indirectly many-body properties of the ground states~\cite{Li:2023msp}. Oxygen-16 is particularly interesting in this respect. It is doubly-magic and thus near spherical in a mean-field description~\cite{Delaroche:2009fa}. However, many-body correlations deform it into an irregular tetrahedral structure with $\alpha$-like clusters at its edges~\cite{Giacalone:2024luz}. Yet, robustly accessing such a pattern in experiments remains an outstanding challenge.

High-energy nuclear collisions open a novel avenue for imaging the ground-state structure of colliding ions~\cite{STAR:2024wgy,Bally:2022vgo}. At large center-of-mass energies, a short passing time between the two nuclei is followed by the formation of a hot and dense quark-gluon plasma (QGP)~\cite{Busza:2018rrf}, whose shape and size in the transverse plane reflect the matter distribution in the colliding nuclei~\cite{Giacalone:2023hwk,Ollitrault:2023wjk,STAR:2024wgy,Jia:2021tzt,Jia:2021qyu}. Observables such as the elliptic flow of outgoing hadrons are in this picture highly sensitive to the details of the spatial distributions of nucleons, including neutron skin and deformation
\cite{Bally:2021qys,PhysRevLett.128.022301,PhysRevLett.127.242301,Nijs:2021kvn,PhysRevLett.131.022301,Ryssens:2023fkv,Giacalone:2023cet,PhysRevLett.125.222301,Chen:2024zwk,PhysRevLett.131.062301,Jia:2021oyt,PhysRevC.106.014906,Wang:2024vjf}. 

With experimental data upcoming from collisions of $^{16}$O nuclei from the BNL Relativistic Heavy Ion Collider (RHIC)~\cite{Huang:2023viw} and the CERN Large Hadron Collider (LHC)~\cite{Brewer:2021kiv}, a pressing question to settle is whether subtle zero-point fluctuations of the $^{16}$O wave function could be revealed by the collider processes. This possibility has triggered a vast body of literature over the years~\cite{Ma:2022dbh}, in assessing the predictions from different models of $^{16}$O encompassing various types of $\alpha$-clustered geometries~\cite{He:2021uko,PhysRevLett.113.032506,YuanyuanWang:2024bxv,Rybczynski:2019adt,PhysRevC.102.054907,PhysRevC.104.L041901,PhysRevLett.112.112501,Shi:2021far,He:2020jzd,Liu:2023gun}. Moreover, understanding the emergence of collective behavior in small system collisions~\cite{CMS:2010ifv,ALICE:2012eyl,ATLAS:2012cix,PHENIX:2018lia,STAR:2022pfn} is a major open issue in the phenomenology of QCD. Therefore, improving our view of the oxygen structure can help more transparently address the nature and intricacies of the final-state dynamics in these processes.

Owing to recent theoretical breakthroughs, in this Letter, we go one step further and perform a comprehensive comparison of the predictions of the existing \abini calculations of the $^{16}$O structure, and study their influence on observables accessible at colliders. We perform high-precision transport simulations that include realistic final-state interaction effects on the experimental observables, thus providing a solid theoretical baseline for the interpretation of future data. Our rationale is straightforward. After briefly recalling the main features of the nuclear models employed in our analysis, we use their predictions to simulate $^{16}$O+$^{16}$O collisions at top BNL RHIC energy. Final-state observables are normalized to the results of simulations employing a 3pF baseline of independent nucleons, given for convenience by the charge density of $^{16}$O. This highlights the impact of nontrivial $NN$ correlations. 

Remarkably, we observe that the final-state observables depend \textit{qualitatively} on the resolution of the spatial correlations present in the sampled nuclear configurations. When the nucleon coordinates are sampled from nuclear Hamiltonians with a low resolution, such as in pion-less EFT, or from clustered and deformed intrinsic one-body distributions, fluctuations in the collective flow of hadrons are enhanced compared to the baseline of results obtained for independent nucleons in a spherical 3pF. However, if harder-scale physics and short-range repulsive effects are more explicitly included in the sampled configurations, then the trend is reversed and the collective flow is suppressed compared to the baseline. This represents our main result, demonstrating the unprecedented opportunities offered by flow measurements in $^{16}$O+$^{16}$O collisions to probe detailed properties of the nuclear ground state. As a bonus, this study enables us to assess the theoretical uncertainty expected on the interpretation of future  $^{16}$O+$^{16}$O data arising from the imperfect knowledge of the oxygen structure.

{\it Nuclear configurations and baseline}-- Recently, configurations of nucleons in the ground state of $^{16}$O have been computed within three different frameworks of \abini nuclear theory, namely:
\begin{enumerate}
    \item Nuclear Lattice Effective Field Theory (NLEFT) simulations~\cite{PhysRevLett.119.222505} employing a minimal pion-less EFT Hamiltonian~\cite{Lu:2018bat}, $L=8$ lattice sites with a spacing $a=1.3155$ fm. The employed resolution scale, $\Lambda=465$ MeV, enables one to account for detailed clustering correlations, which is a low-resolution phenomenon. Repulsive effects acting on scales of a fermi or so are instead not included in this approach.  The configurations used in this work are taken from Ref.~\cite{Summerfield:2021oex}, and implement a Gaussian smearing of width 0.84 fm (corresponding to the proton charge radius \cite{Lin:2021xrc}) at each lattice site.
    \item Variational Monte Carlo - Auxiliary Field Diffusion Monte Carlo (VMC for brevity) simulations based on an N$^2$LO chiral EFT Hamiltonian~\cite{Lonardoni:2018nob}, with a resolution scale $\Lambda=500$ MeV which enables one to resolve short-range repulsive effects therein encoded. The configurations we utilize are taken from Ref.~\cite{Lim:2018huo}.
    \item \abini Projected Generator Coordinate Method (PGCM) calculations~\cite{Frosini:2021fjf,Frosini:2021sxj,Frosini:2021ddm} based on an N$^3$LO chiral EFT Hamiltonian and a resolution scale of 500 MeV. In this approach, nucleon positions are not sampled directly according to the ground state of the Hamiltonian, but rather from a clustered and deformed mean-field density~\cite{Bally:2020kjr,Bally:2024loa}. This procedure implies that the sampled coordinates will effectively capture low-resolution phenomena in the wave function (clustering and deformation). However, nucleons are sampled independently from the randomly-oriented density without adding short-range correlations~\cite{Giacalone:2024luz}. Therefore, even though the N$^{3}$LO Hamiltonian contains in principle more information than than the N$^2$LO Hamiltonian employed in the VMC computations, high-resolution features are disregarded in the \textit{ab initio} PGCM approach due to the simplified sampling method.
\end{enumerate}

\begin{figure*}[t]
\centering
\includegraphics[width=.9\linewidth]{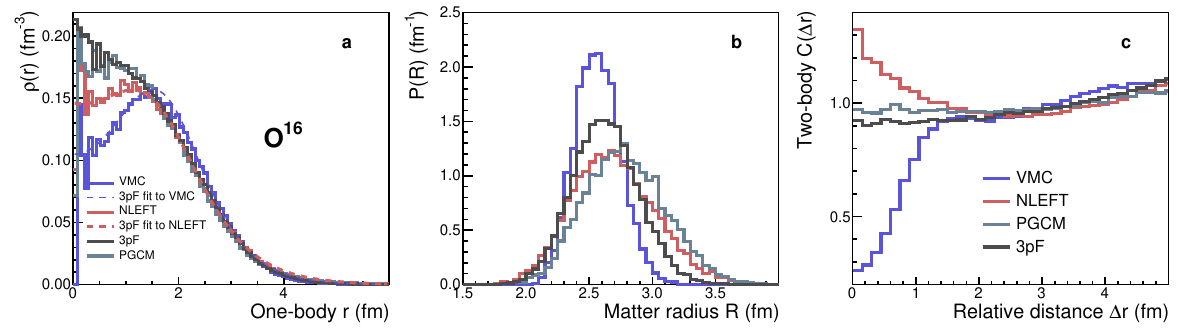}
\caption{\label{fig1}$\bold{a}$, The nuclear one-body density, $\rho(r)$, is depicted for various models of the $^{16}$O. We note that the 3pF baseline given by the charge density does not have the expected flat shape towards the center of the nucleus. The reason is that the displayed curve is obtained from sampled configurations that are re-centered, to ensure an apples-to-apples comparisons with the VMC and NLEFT evaluations. The same re-centering is applied to the PGCM nucleons. $\bold{b}$, The probability density of the matter rms radius, $P(R)$. $\bold{c}$, Two-body correlation function, $C(\Delta r)$, where $\Delta r$ is the inter-nucleon separation.}
\end{figure*}
In Fig.~\ref{fig1} we display the main features of the radial distributions of the sampled nucleons in the aforementioned theories. For the one-body density in Fig.~\ref{fig1}a, we note an ordering between the results, with the VMC curve showing the strongest density depletion at small radii. The central depletion is a consequence of nuclear clustering: if nucleons tend to sit in clusters located at $r>0$, then the density is depleted at the center. The observed ordering well reflects the expectations from the implemented correlations. The VMC configurations appear to present the strongest clustering patterns. The figure shows as well the results from a 3pF fit to the measured charge density of $^{16}$O, for which $R=2.608$ fm, $a_0=0.513$ fm, and $w=-0.051$.

Figure~\ref{fig1}b shows instead the distribution of matter rms radii, $R =\sqrt{\frac{1}{16} \sum_i (x_i^2+y_i^2+z_i^2)}$, with $i=1,\ldots,16$, in the three models. The nucleons have been recentered such that $\lr{x}=\lr{y}=\lr{z}=0$ in each nucleus. We note that the NLEFT and PGCM results align well with the results obtained by sampling nucleons independently from the 3pF density. This is expected from the lower resolution of the NLEFT Hamiltonian, as well as from the fact that the PGCM coordinates come essentially from an independent sampling, albeit with nontrivial angular correlations that are not important for this plot. The VMC results are, on the other hand, dramatically different. They show a strong departure from the mean-field baseline, which is a consequence of stronger correlations probed by the VMC sampling method which takes into account higher-resolution features of the Hamiltonian.

Finally,  Fig.~\ref{fig1}c displays the distance-distance correlation among nucleon pairs, commonly quantified via a normalized two-body distribution,
\begin{equation}
    C(\Delta r) = \frac{ \int d^3 {\bf s}~ f \left ( {\bf s} + \frac{{\bf r}}{2} , {\bf s} - \frac{{\bf r}}{2} \right ) }{ \int d^3{\bf s} ~f\left ( {\bf s} + \frac{{\bf r}}{2} \right ) f \left ( {\bf s} - \frac{{\bf r}}{2} \right )  }, \hspace{10pt}\Delta r = |{\bf r}|,
\end{equation}
where ${\bf s}$ is the center of mass of a nucleon-nucleon pair, while ${\bf r}$ is the relative distance vector. The two-body distribution in the numerator is calculated from pairs from the same batch of $A=16$ coordinates, and then averaging over configurations. The nucleon pairs in the denominator are taken from different parent nuclei (\textit{mixed event}), erasing thus any genuine two-body effects~\cite{Rybczynski:2019adt}. In Fig.~\ref{fig1}c, at short distances, the positive correlation observed in the NLEFT results is a remnant of the underlying theory being discretized on a lattice. The notable feature in this plot is the repulsive correlation present in the VMC evaluations, reflecting hard-scale physics included in the chiral interaction and probed by the sampling.

{\it Dynamical transport calculations}-- We move on then to the dynamical simulations of $^{16}$O+$^{16}$O collisions. The event-by-event simulations are performed by means of the A Multi-Phase Transport (AMPT) framework~\cite{Lin:2004en}, which provides a reasonable description of multiplicity distributions and flow signatures in both small and large collision systems at RHIC and the LHC~\cite{Ma:2014pva,PhysRevLett.113.252301}. The working principles of the model are recalled in the Supplemental Material (SM). We perform simulations starting from the \abini nuclear configurations, from configurations sampled according to the 3pF charge density baseline, as well as from configurations sampled from 3pF distributions that are fitted to the NLEFT and VMC one-body densities of Fig.~\ref{fig1}a (we obtain $R=1.84$ fm, $a_0=0.46$ fm, and $w=1.76$ for the VMC fit and $R=1.65$ fm, $a_0=0.53$ fm, and $w=0.87$ for the NLEFT fit). We compute the final-state observables as a function of the collision centrality, defined from the charge multiplicity at midrapidity (with 0\% corresponding to the limit of fully-overlapping nuclei). We focus on quantities that probe fluctuations and correlations in the interaction region, namely, the radial and elliptic flow of produced hadrons. We compute the mean transverse momentum, $\langle p_{T}\rangle$, the mean squared elliptic flow, $v_2^2$, the variance and the skewness of the mean transverse momentum fluctuations, respectively, $\langle (\delta p_T)^2 \rangle$ and $\langle (\delta p_T)^3 \rangle$, as well as the covariance of the elliptic flow and the mean transverse momentum, $\langle v_2^2\delta p_T \rangle$, for events taken from narrow centrality classes. The operational definition of these observables in terms of final-state particle distributions is presented in the SM [Eqs.~(\ref{eq2}-\ref{eq4})].

\begin{figure*}[t]
\centering
\includegraphics[width=0.95\linewidth]{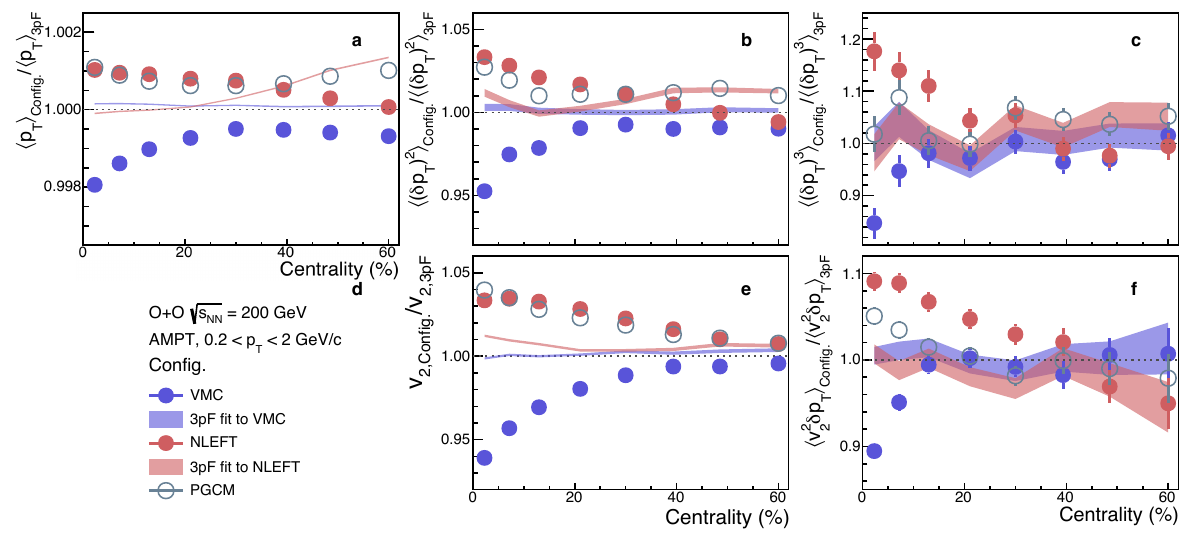}
\caption{\label{fig2} Bulk observables computed in \oo{} collisions at $\sqrt{s_{NN}} =$ 200 GeV for different models of the oxygen structure are normalized to those obtained starting from the charge density 3pF parametrization of the nucleus. The ratios are shown as a function of the collision centrality. $\bold{a}$, the mean transverse momentum, $\langle p_T\rangle$. Two-particle correlation observables are instead the variance of mean transverse momentum fluctuation $\left\langle(\delta p_{\mathrm{T}})^2\right\rangle$, $\bold{b}$, and the rms elliptic flow $v_2$, $\bold{e}$, respectively. Three-particle correlation observables are the skewness of the mean transverse momentum fluctuation $\left\langle(\delta p_{\mathrm{T}})^3\right\rangle$, $\bold{c}$, and the covariance $\left\langle v_2^2 \delta p_T\right\rangle$, $\bold{f}$, respectively. The specific configuration labels are shown in panel $\bold{d}$.}
\end{figure*}

The final-state observables obtained in the different calculations are normalized to the results obtained starting from the 3pF charge density. Hence, if the ratio is unity the result is essentially consistent with an independent sampling of nucleons from the charge density profile, whereas the presence of $NN$ correlations will cause the ratios to deviate from the baseline. Our results are displayed in Fig.~\ref{fig2}. We note a rather universal trend for all observables: compared to the baseline, $NN$ correlations tend to enhance the fluctuations of the collective flow for the NLEFT and PGCM scenarios, while they deplete the observables for the VMC input. As expected, these effects are more prominent in central collisions, where the entire geometry of the colliding ions is resolved by the interactions. In addition, as argued in the SM, the results are robust against non-flow correlations and viscosity changes. We conclude that these observables are effectively probing the details of the spatial correlations carried by the nuclear configurations, as dictated by the resolution of the underlying Hamiltonians and by the quality of the sampling of the many-body wave function. The good consistency of the the NLEFT and the PGCM results indicates that the low-resolution phenomena captured within NLEFT simulations based on the SU(4)-symmetric pion-less EFT interaction can essentially be reproduced by sampling nucleons according to randomly-oriented deformed and clustered densities predicted by the \textit{ab initio} PGCM approach.  Assuming, then, that the NLEFT and VMC many-body solutions and sampling methods for the ground state are equally good, one would conclude that the spread in results observed in Fig.~\ref{fig2} arises from the stronger short-distance correlations present in the VMC configurations. 

\begin{figure*}[t]
\centering
\includegraphics[width=0.70\linewidth]{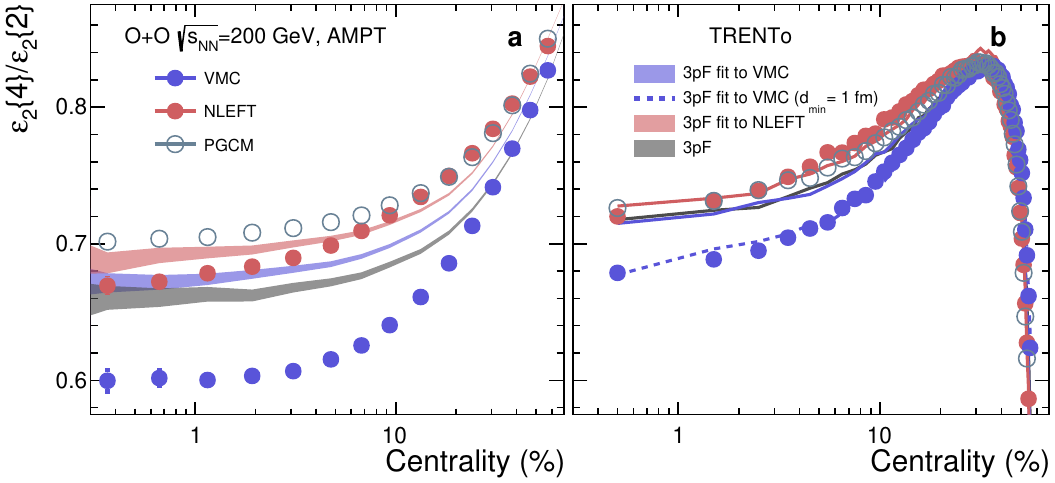}
\caption{\label{fig3}Relative eccentricity fluctuation, $\varepsilon_2\{4\}/\varepsilon_2\{2\}$, plotted as a function of the centrality percentile in $^{16} \mathrm{O}+^{16} \mathrm{O}$ collisions at $\sqrt{s_{NN}} =$ 200 GeV, for different models of the oxygen structure. \textbf{a}. Results from AMPT Glauber initial conditions. \textbf{b}. Results from \trento{} initial conditions.}
\label{fig:3}
\end{figure*} 

Another important aspect of our study is the comparison with the results obtained starting from fitted 3pF densities. In central collisions, such results do not show a significant departure from the baseline, although the underlying radial one-particle distributions are fairly different (see Fig.~\ref{fig1}a). This means that the relevant correlations affecting the observables displayed in Fig.~\ref{fig2} are angular, rather than involving only radial information. 

Before concluding, we connect this work with the recent analysis of preliminary $^{16}$O+$^{16}$O data performed by the STAR collaboration~\cite{Huang:2023viw}. The observable under study is the relative fluctuations of elliptic flow~\cite{Giacalone:2017uqx,Bozek:2014cva}, given by the ratio $v_2\{4\}/v_2\{2\}$ of the fourth-order cumulant of the elliptic flow distribution, $v_2\{4\}$, to the second-order one, $v_2\{2\}$ (see the SM for operational definitions). Obtaining a four-particle correlation observable with adequate precision would require several simulations with event statistics comparable to those recorded experimentally (close to half-billion simulations), which is beyond our current possibilities. Therefore, we only give an estimate of this quantity, assuming a proportionality relation between $v_2$ and the participant-plane eccentricity, $\varepsilon_2$ in a given centrality class, $v_2 \propto \varepsilon_2$. This implies that approximately $\varepsilon_2\{4\}/\varepsilon_2\{2\} \approx v_2\{4\}/v_2\{2\}$, which we can obtain with a higher precision from initial-state calculations. 

In Fig.~\ref{fig3}a, we show our results from the AMPT Glauber initial conditions. Different nuclear structure inputs yield results that agree on a quantitative level. The only outlier is the result obtained from the VMC configurations. To assess the robustness of this finding, we repeat the calculation of the eccentricity fluctuations by using this time the popular \trento{} model of initial conditions~\cite{Moreland:2014oya}, with the same parameter setup as in Ref.~\cite{Bally:2023dxi}, where the eccentricity is defined in each event from the created entropy profiles. The results are shown in Fig.~\ref{fig:3}b. We note the same pattern as in the other panel, with all calculations agreeing quantitatively except for the VMC results. To shed more light on this behavior, we repeat the \trento{} calculation that starts with a 3pF density fitted to the VMC result, by implementing this time a short-range repulsion among nucleons, namely, we forbid two sampled nucleons to stay at a distance closer than $d_{\rm min}=1$ fm. The calculation including this effect is shown as a dashed line in Fig.~\ref{fig:3}b. It matches nicely the result obtained from the full VMC computation, corroborating the point that the $v_2\{4\}/v_2\{2\}$ observable will indeed enable us to place experimental constraints on the structure of $^{16}$O in its ground state.

Finally, it is interesting to note that the eccentricity fluctuation obtained from the participant nucleons in the Glauber AMPT initial condition calculation leads to a value of $\varepsilon_2\{4\}/\varepsilon_2\{2\}$, which is in good agreement with the recent STAR measurements~\cite{Huang:2023viw}. The \trento{} results are instead significantly higher. This means that a non-linear hydrodynamic response between the ellitpic flow and the eccentricity will be needed to match hydrodynamic calculations to the experimental data. Such non-linear effects seem to be present in hydrodynamic results for $^{208}$Pb+$^{16}$O collisions \cite{Giacalone:2024ixe}, but have yet to be tested in $^{16}$O+$^{16}$O collisions.
 
{\it Conclusion and outlook}-- Exciting times are ahead with the advent of $^{16}$O+$^{16}$O collisions at RHIC and the LHC. Combining the most advanced evaluations of the ground-state structure of $^{16}$O with high-precision transport simulations of the collision processes, our results suggest that bulk observables in $^{16}$O+$^{16}$O collisions will offer a high-resolution view of the inner oxygen structure and the influence of many-body correlations. A plethora of observables can be used for this purpose, as indicated by the results in Fig.~\ref{fig2}. Such a possibility, combined in the future with advanced Bayesian analysis \cite{Paquet:2023rfd}, will effectively permit us to directly link the properties of the nuclear Hamiltonians to the collider data, and further infer the main impacts of low-energy constants in the chiral interaction on the collective flow~\cite{Sun:2024iht,Ekstrom:2023nhc}. We expect thus $^{16}$O+$^{16}$O collisions to foster a new program of interdisciplinary QCD studies.

Conversely, the present analysis helps in the preparation of the hot QCD program envisaged for the $^{16}$O+$^{16}$O runs with two main expectations~\cite{Brewer:2021kiv}. One pertains to the observation of jet quenching and in a small system \cite{Huss:2020dwe}. Our results suggest that quantities probing the size of the interaction region, such as $\langle \pT \rangle$ \cite{Schenke:2020uqq,Giacalone:2020dln}, come with a small 0.1\% uncertainty from the input nuclear model. Therefore, the size of the medium expected to be traversed on average by the hard probes is expected to be well constrained. Secondly, $^{16}$O+$^{16}$O collisions may permit us to better unveil the origin of the collectivity \cite{Nagle:2018nvi,Noronha:2024dtq,Huang:2019tgz,Grosse2024bwr}, as well as the detail longitudinal decorrelation dynamics in small systems \cite{Zhang:2024bcb,STAR:2023wmd}. From Fig.~\ref{fig2}, we expect that characterizations of the collective dynamics to ascertain the presence of a hydrodynamic QGP-like medium will be hindered by considerable uncertainties, of order 10-20\%, associated with the nuclear inputs. These can be reduced by looking at collision centralities higher than 40\%, although at the cost of less produced particles. Alternatively, combining $^{16}$O+$^{16}$O collisions with other collision systems having near equal participant nucleon numbers will largely reduce the theory uncertainty coming from the nuclear modeling, in particular from the modeling of the short-range correlations~\cite{Huang:2019tgz,Giacalone:2024luz}.

{\it Acknowledgements }-- The authors are indebted to Somadutta Bhatta for his helpful comments. We are grateful to Guoliang Ma, Dean Lee, You Zhou, Simin Wang, Xinli Zhao, Benjamin Bally, Govert Nijs, Wilke van der Schee, and Vittorio Som\`a for useful discussions. C. Zhang, J. Chen, and Y. Ma are founded by the National Key Research and Development Program of China under Contract No. re2022YFA1604900, by the National Natural Science Foundation of China under Contracts Nos. 12025501 and 12147101. S. Huang and J. Jia are supported by DOE Research Grant Number DE-SC0024602. G. Giacalone is funded by the Deutsche Forschungsgemeinschaft (DFG, German Research Foundation) - Project ID 273811115 - SFB 1225 ISOQUANT, and under Germany's Excellence Strategy EXC2181/1-390900948 (the Heidelberg STRUCTURES Excellence Cluster).

\bibliography{ref}{}
\bibliographystyle{apsrev4-1}
\section{Appendix}
\paragraph{Calculations of observables}-- The 2nd and 4th order cumulants of the distribution of $\varepsilon_2$ vector are defined as, 
\begin{equation}\label{eq1}\begin{split}
& \varepsilon_2\{2\}^2=\left\langle \varepsilon_2^2\right\rangle \\
& \varepsilon_2\{4\}^4=2\left\langle \varepsilon_2^2\right\rangle^2-\left\langle \varepsilon_2^4\right\rangle
\end{split}
\end{equation}

The elliptic flow $\left\langle v_2^2\right\rangle$ is calculated based on a multi-particle correlation framework prescribed in Refs.~\cite{Jia:2017hbm,Zhang:2021phk,Bhatta:2021qfk} as, 
\begin{equation}\label{eq2}\begin{split}
\left\langle v_2^2\right\rangle& =\left\langle\frac{\sum_{i \neq j} w_i w_j \cos [2\left(\phi_i-\phi_j\right)]}{\sum_{i \neq j} w_i w_j}\right\rangle
\end{split}
\end{equation}
Mean transverse momentum in each event $\left[p_{\mathrm{T}}\right]$ and its high-order flucutations, variance and skenwess $\left\langle\left(\delta p_{\mathrm{T}}\right)^n\right\rangle$ for $n=2,3$, respectively, are given by:
\small\begin{equation}\label{eq3}\begin{split}
\left[p_{\mathrm{T}}\right]&=\frac{\sum_i w_i p_{\mathrm{T}, i}}{\sum_i w_i},\left\langle\left\langle p_{\mathrm{T}}\right\rangle\right\rangle \equiv\left\langle\left[p_{\mathrm{T}}\right]\right\rangle_{\mathrm{evt}}\\
\left\langle\left(\delta p_{\mathrm{T}}\right)^2\right\rangle&=\frac{\sum_{i \neq j} w_{i} w_{j}\left(\delta p_{\mathrm{T}, i}\right)\left(\delta p_{\mathrm{T}, j}\right)}{\sum_{i\neq j} w_{i} w_{j}}\\
\left\langle\left(\delta p_{\mathrm{T}}\right)^3\right\rangle&=\frac{\begin{split}
\sum_{i \neq j \neq k} w_iw_jw_k\left(\delta p_{\mathrm{T}, i}\right)\left(\delta p_{\mathrm{T}, j}\right)\left(\delta p_{\mathrm{T}, k}\right)
\end{split}
}{\sum_{i \neq j \neq k} w_iw_j w_k}
\end{split}
\end{equation}\normalsize
where $\delta p_{\mathrm{T}, i} = p_{\mathrm{T}, i}-\left\langle\left\langle\mathrm{p}_{\mathrm{T}}\right\rangle\right\rangle$. The track-wise weight $w_{i,j,k}$ accounts for the experimental detector effect set to be 1 in the model simulation. Correlations between elliptic flow and mean transverse momentum, $\left\langle v_2^2 \delta p_{\mathrm{T}}\right\rangle$~\cite{Bozek:2016yoj} is expressed as  
\begin{equation}\label{eq4}\begin{split}
\left\langle v_2^2 \delta p_{\mathrm{T}}\right\rangle&=\left\langle\frac{\sum_{i \neq j \neq k} w_i w_j w_k e^{i2(\phi_i-\phi_j)}\left(\delta p_{\mathrm{T}, k}\right)}{\sum_{i \neq j \neq k} w_i w_j w_k}\right\rangle_{\mathrm{evt}}
\end{split}
\end{equation}

The averages are performed by looping over particles to obtain all unique multiples within a single event and then averaged over an ensemble of events within fixed centrality intervals. The $\left\langle v_2^2\right\rangle$ is calculated via the two-subevent method, where particle $i$ and $j$ are selected from different pseudorapidity ranges of $-2<\eta <-0.5$ and $0.5 < \eta < 2$, respectively, for suppressing the short-range correlations, commonly referred to as ``non-flow", arising from resonance decays and jets~\cite{STAR:2022pfn}. The skewness $\left\langle\left(\delta p_{\mathrm{T}}\right)^3\right\rangle$ and covariance $\left\langle v_2^2 \delta p_{\mathrm{T}}\right\rangle$ are calculated by averaging over all triplets labeled by particles indices $i$, $j$, and $k$. The standard cumulant framework is used to obtain the results instead of directly calculating all triplets.

\paragraph{Model step and simulations} -- AMPT model~\cite{Lin:2004en} implemented four main phases, fluctuating initial conditions from the HIJING model, elastic parton cascade simulated by the ZPC model, quark coalescence model for hadronization, and hadronic re-scattering based on ART model. Utilizing the version v2.26t9 in string melting mode at $\sqrt{s_{NN}} = $ 200 GeV, with a partonic cross section of 3.0 $m$b, provides a reasonable description of RHIC data~\cite{PhysRevC.84.014903}. In the current study, generic \oo collisions are initiated incorporating five configurations to sample the nucleon spatial distributions within the \abini approaches and 3pF by generating 300 million mininum-bias events for each case. The final-state hadrons within $0.2<p_{\mathrm{T}}<2$ $\mathrm{GeV}/c$ and $|\eta| <$ 2 are analyzed to achieve the best statistical precision. The centrality is defined by charged hadron multiplicity measured at $|\eta|<0.5$, following STAR experiment's ``Refmul" centrality definition.

\begin{figure}[t]
\centering
\includegraphics[width=0.75\linewidth]{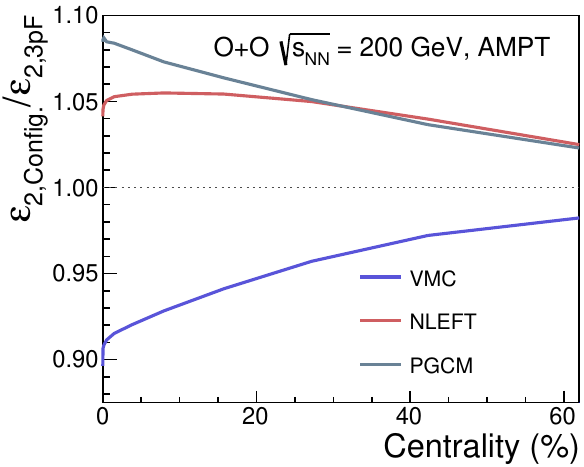}
\caption{\label{fig4}The centrality dependence of $\varepsilon_2$ ratios for the ratios of VMC, NLEFT and PGCM to the traditional 3pF configurations in initial state $^{16} \mathrm{O}+^{16} \mathrm{O}$ collisions at $\sqrt{s_{NN}} =$ 200 GeV.}
\end{figure}
\begin{figure}[t]
\centering
\includegraphics[width=0.75\linewidth]{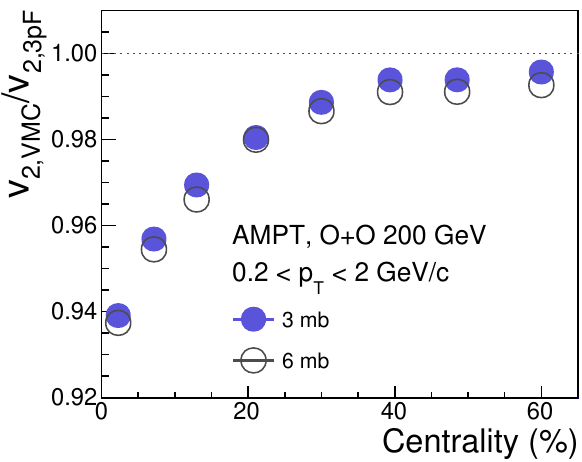}
\caption{\label{fig5}The centrality dependence of $v_2$ ratios for two different values of partonic cross sections 3 $m$b and 6 $m$b for VMC to the traditional 3pF configuration, using hadrons in $|\eta|<2$ and $0.2<p_{\mathrm{T}}<2$ $\mathrm{GeV}/c$ in $^{16} \mathrm{O}+^{16} \mathrm{O}$ collisions at $\sqrt{s_{NN}} =$ 200 GeV.}
\end{figure}

In  Fig.~\ref{fig4}, We observe that the $\varepsilon_2$ model exhibits similar, but stronger ordering when compared to the final state $v_2$ behaviors in Fig.~\ref{fig2}. 
\paragraph{Final state interactions}-- 
Following our previous work~\cite{PhysRevC.106.L031901}, the different combinations of QCD coupling constant $\alpha_s$, and screening mass $\mu$ give the different partonic cross sections, and also the different specific shear viscosity $\eta/s$. The role of the final state effects is studied reasonably varying the $\eta/s$ from 0.232 to 0.156, up and down by 30\%, by changing the partonic cross-section from 3 mb to 6 mb. These choices significantly change the predicted $v_2$, yet the $v_2$ ratios remain unchanged as shown in Fig.~\ref{fig5}, implying that the ratios of elliptic flow are insensitive to the medium properties in the final state.

\begin{figure}[t]
\centering
\includegraphics[width=0.75\linewidth]{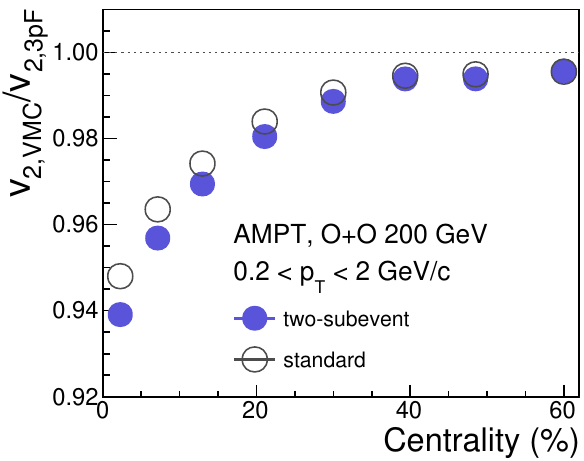}
\caption{\label{fig6}The centrality dependence of $v_2$ ratios for the standard and two-subevent method for VMC to the traditional 3pF configuration, using hadrons in $|\eta|<2$ and $0.2<p_{\mathrm{T}}<2$ $\mathrm{GeV}/c$ in $^{16} \mathrm{O}+^{16} \mathrm{O}$ collisions at $\sqrt{s_{NN}} =$ 200 GeV.}
\end{figure}

\begin{figure*}[t]
\centering
\includegraphics[width=0.85\linewidth]{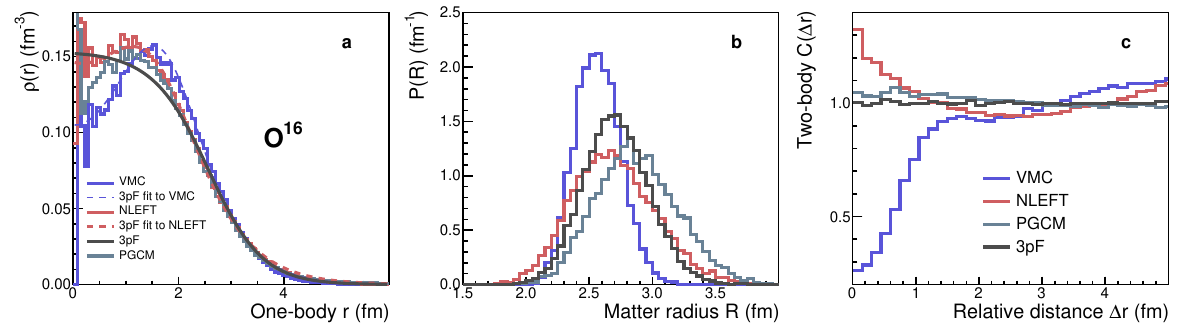}
\caption{\label{fig7} Same as Fig.~\ref{fig1} except that the PGCM and 3pF cuves are obtained without re-centering the configurations.}
\end{figure*}

\paragraph{Nonflow effects}--
The possible contamination from ``nonflow" on anisotropic flow in small systems cannot be fully eliminated. However, the nonflow should be largely diluted in ratios between different configurations within the same centrality intervals in the same collision system. Therefore, any deviation of the ratio from unity is still a strong indication of the role of nucleonic clustering in the initial state. Figure~\ref{fig6} demonstrates that the ``nonflow" effect is less than 1\% which is rather small compared with the nuclear structure effect.

\paragraph{Recentering effects} -- For completeness, we show in Fig.~\ref{fig7} the same distributions presented in Fig.~\ref{fig1}, albeit without the re-centering the sampled PGCM and 3pF nucleons. For the 3pF baseline, we recover the expected flat behavior at small radii for the one-body density (panel \textbf{a}). Beside that, we note in Fig.~\ref{fig7}c that switching off re-centering eliminates the slight positive slope of the two-body correlation that was observed in the PGCM results and the 3pF baseline. This suggests that the features exhibited by the NLEFT and VMC nucleons might be an artifact of re-centering.

\end{document}